\documentclass[12pt]{iopart}
\usepackage{graphicx}
\usepackage{iopams}
\usepackage{textcomp}
\usepackage{xcolor}
\usepackage{ulem}

\begin{document}
%\JPCM 

\title[ESR on YbCl$_3$]{Electron spin resonance study on the 4$f$ honeycomb quantum magnet YbCl$_3$}

\author{J\"org Sichelschmidt$^1$, Ellen H\"au{\ss}ler$^2$, Ekaterina Vinokurova$^{3,4}$, M. Baenitz$^1$ and Thomas Doert$^2$}

\address{$^1$ Max Planck Institute for Chemical Physics of Solids, Dresden, Germany}
\address{$^2$ TU Dresden, Dept. of Chemistry and Food Chemistry, Dresden, Germany}
\address{$^3$ TU Dresden, Institut f\"ur Festk\"orper- und Materialphysik, 01062 Dresden, Germany}
\address{$^4$ Leibniz IFW Dresden, Institute of Solid State Research, 01069 Dresden, Germany}
\ead{Sichelschmidt@cpfs.mpg.de}

\begin{abstract}
The local magnetic properties of Yb$^{3+}$ in the layered honeycomb material YbCl$_{3}$ were investigated by electron spin resonance on single crystals. For in-plane and out-of-plane field orientations the $g$-factor shows a clear anisotropy ($g_\|=2.97(8)$ and $g_\bot =1.53(4)$), whereas the low temperature exchange coupling and the spin relaxation display a rather isotropic character. At elevated temperatures the contribution of the first excited crystal field level ($21\pm2$~meV) dominates the spin relaxation.
\end{abstract}

\section{Introduction}

Yb-containing quantum magnets in general but planar triangular or honeycomb lattices in particular are experiencing a resurgence of strong interest due to the emerging spin-liquid physics towards lowest temperatures. Due to spin-orbit entanglement, the exchange interactions are complex and bond-dependent, leading to strong frustration between spin-1/2 Yb Kramers ions.  For the Yb-based planar triangular lattice spin liquids NaYbCh$_{2}$ (Ch:O,S,Se) we have shown the electron spin resonance (ESR) of Yb to be a powerful local probe to study the local static and dynamic magnetization \cite{sichelschmidt19a,sichelschmidt20a,schmidt21a}. Due to its high resolution, the ESR method is particularly suitable for the investigation of the magnetic properties of very small single crystals, which is a clear advantage over conventional methods (SQUID magnetometry) \cite{haussler22a}. 

Yb-based honeycomb lattices have been little studied so far.  Ir$^{4+}$ honeycomb lattices and Ru$^{3+}$ honeycomb lattices, on the other hand, have been the focus of research in recent years. Here, the complex exchange can be described quite well with the so-called Kitaev model. This model predicts a number of magnetically ordered states including the non-ordered spin liquid state (see review \cite{trebst22a} and references in there). 

YbCl$_{3}$ is a proposed candidate material for Kitaev physics on a honeycomb lattice containing edge-sharing octahedra \cite{rau18a,xing20b,templeton54a} and is isostructural to the quantum spin liquid (QSL) candidate material $\alpha$-RuCl$_{3}$. It exhibits short-range magnetic order below 1.2~K and N\'eel-type antiferromagnetic long-range order below 0.6~K \cite{xing20b,hao20a}. Strong quantum fluctuations were inferred from a reduced ordered moment and a tiny magnetic entropy release.
It is discussed in terms of a model specified on anisotropic spin exchange interactions of rare-earth magnets on an unfrustrated honeycomb lattice \cite{luo20a}. This model also discusses the magnetic anisotropy as the main source of broadening of Yb$^{3+}$-ESR in YbCl$_{3}$.

Here, we present the Yb$^{3+}$-ESR in YbCl$_{3}$ which proofs a strong $g$ factor anisotropy, a contrasting low-temperature behavior between in-plane and out-of-plane line broadening, and a small anisotropy in Weiss temperatures which points towards a weak exchange anisotropy.

\section{Experimental}
\subsection{Electron Spin Resonance}
We performed  our ESR investigations as previously described in more detail \cite{sichelschmidt19a}. We used a magnetic microwave field $b_{mw}$ at $\nu $=9.4 GHz and transversally aligned to an external magnetic field $\mu_0H=B$. The ESR signal consists of the derivative of the absorbed microwave power $dP/dB$ and is measured for temperatures between 2.4 and 65~K.
A Lorentzian line shape fitting yielded the parameters linewidth $\Delta B$, determined by the spin-probe relaxation, and resonance field $B_{\rm res}$, determined by the effective $g$-factor $(g=h\nu/\mu_{\rm B}B_{\rm res}$) and internal fields.
The integrated ESR absorption $I_{A}$ was calculated using the line amplitude and linewidth \cite{gruner10a}. It is a measure of the intensity $I_{ESR}=I_{A}g$ being determined by the static spin-probe magnetic susceptibility $\chi_{\rm ESR}$ along the microwave magnetic field.

\subsection{Sample preparation and characterisation}

\begin{figure}
\centering
      \includegraphics[width=0.8\linewidth]{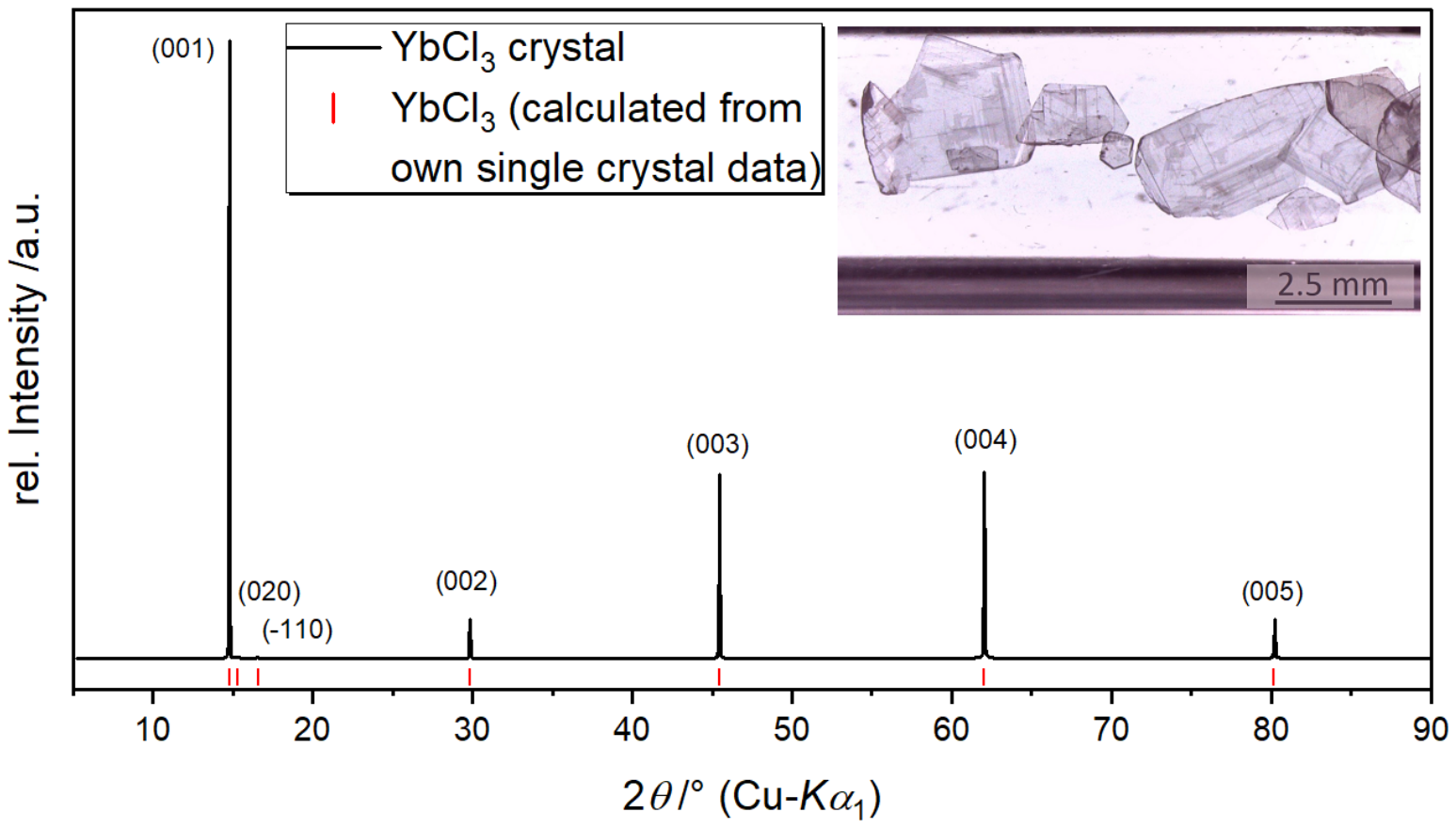}
      \caption{Powder X-ray diffraction pattern measured on a YbCl$_3$ single crystal under inert gas. Note that mainly 00l reflections are visible due to the strong preferred orientation under this measurement condition. Reflection positions were calculated from own single crystal diffraction data.
      The inset shows the typical size of crystals in the sink side of the ampule.}
      \label{fig:YbCl3_PXRD}
\end{figure}

As YbCl$_3$ is highly water sensitive, all operations were carried out in argon-filled glove boxes with O$_2$ and H$_2$O content less than 0.1 ppm. Special care was also taken to store and transport YbCl$_3$  samples.
Crystals were grown by chemical vapour transport from purchased YbCl$_3$ powder (anhydrous, 99.9~\%, ChemPUR) with AlCl$_3$ (anhydrous, 98~\%, Sigma Aldrich) as the transport agent in molar ratios of 5:1.
Both powders were mixed in an agate mortar and filled into the quartz ampule, which was sealed under vacuum. The ampule was placed in a two zone furnace with the source side at 1073~K and the sink side at   873~K. After five days, thin transparent colourless crystals with up to 5 mm edge length were found in the sink side (see inset of Fig.~\ref{fig:YbCl3_PXRD}). Smaller crystals and white powder with intergrown crystals were found in the middle and in the source side of the ampule, respectively. 

\begin{figure}
\centering
      \includegraphics[width=0.6\linewidth]{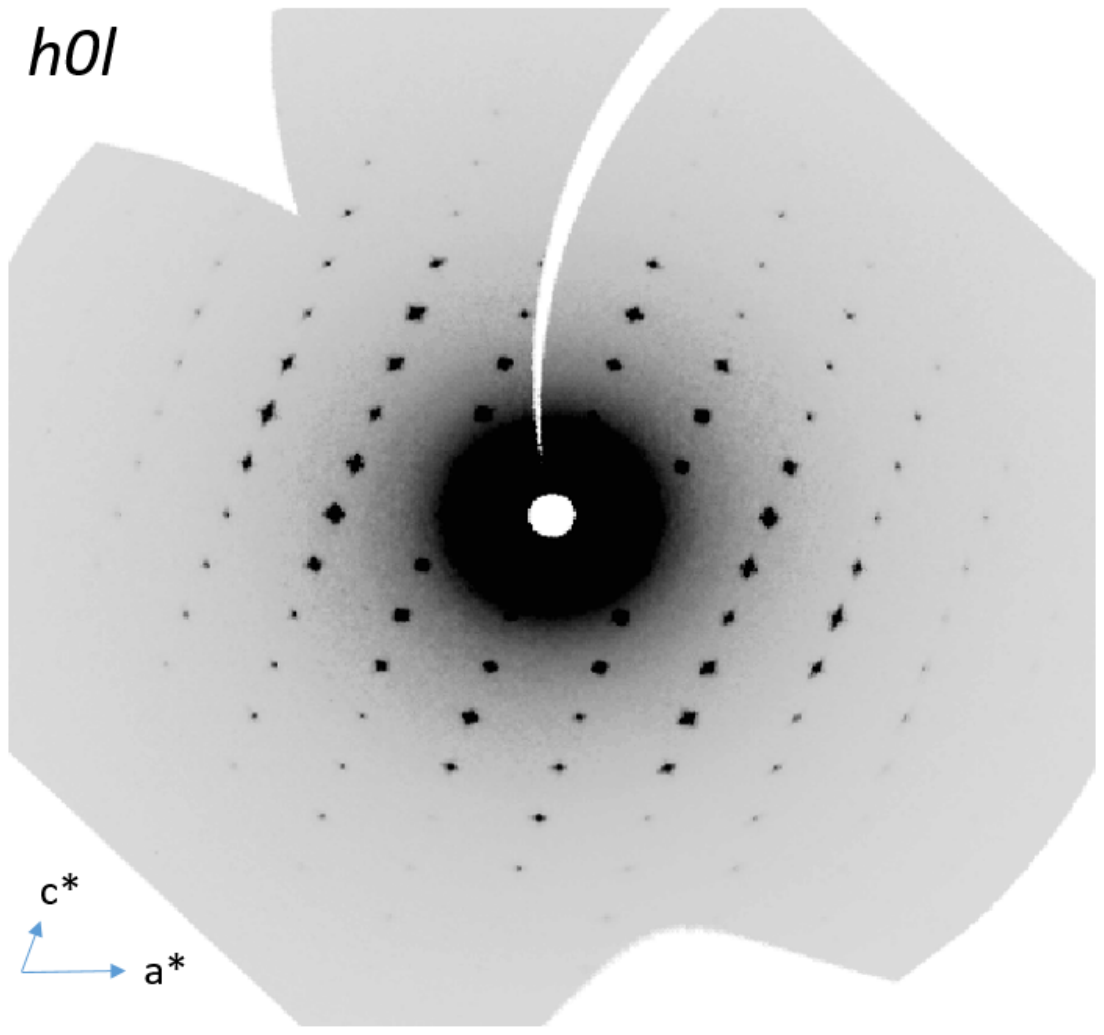}
      \caption{Reconstructed reciprocal space layer.}
      \label{fig:h0l}
\end{figure}

Powder X-Ray Diffraction (PXRD) pattern was collected on a single crystal of YbCl$_3$ in an inert gas sample holder. The data was recorded at room temperature on PANalytical X'Pert Pro diffractometer equipped with a curved Ge(111) monochromator with Cu-K$\alpha_1$-radiation ($\lambda=1.54056$ \AA) in Bragg--Brentano geometry in the range $5^\circ\leq2\theta\leq90^\circ$. As the pattern was collected from a single crystal, it is highly textured with mostly only reflections of $(00l)$-series visible,  Fig.~\ref{fig:YbCl3_PXRD}. All visible reflections can be indexed with a monoclinic $C2/m$ space group in accordance with literature \cite{xing20b} and the theoretical pattern calculated from our single-crystal data, see below. No reflections of impurity phases were observed. 
 
Single-Crystal X-Ray Diffraction (SCXRD) was performed on STOE IPDS II diffractometer with Ag-K$\alpha$ ($\lambda=0.56087$ \AA) radiation at ambient temperature. A suitable crystal was selected and sealed in a glass capillary inside a glove box under inert gas atmosphere. Data collection, processing, integration, and numerical absorption correction were performed using the X-area software packages X-SHAPE and X-RED32 \cite{X-Shape,X-RED32}. The structure was solved with the SHELXT \cite{shelxt} structure solution program and refined using Olex2 \cite{Olex2} with the SHELXL \cite{shelxl} refinement package. Crystallographic data and experimental details are summarized in Table~\ref{crystaldata} and Table~\ref{atomic positions}. Our data confirm YbCl$_3$ to crystallize in the monoclinic space group $C2/m$ with lattice parameters $a=6.725(3)$\AA, $b = $ 11.591(5) \AA, $c = $ 6.361(3) \AA, $\beta = 110.33(4)^{\circ}$  in agreement with the literature \cite{xing20b}. No deviation from the ideal composition was found within experimental limits. The crystal structure data are deposited in the Inorganic Crystal Structure Database (ICSD) at FIZ-Karlsruhe and can be obtained free of charge by quoting the deposition number CSD-2250749.

\begin{table}
\scriptsize
\caption{\label{crystaldata}Crystal data, structure refinement and experimental details for YbCl$_3$ }
\begin{tabular*}{\textwidth}{@{}l*{15}{@{\extracolsep{0pt plus 12pt}}l}}
\br
Crystal system	& monoclinic\\
Space group & $C2/m$\\
$a$, \AA	 &6.725(3)\\
$b$, \AA	 &11.591(5)\\
$c$, \AA	 &6.361(3)\\
$\beta$, $^{\circ}$	 &110.33(4)\\
$V$, \AA$^3$  &464.9(4)\\
Z & 4\\
2$\theta$ range, $^{\circ}$ & from 5.39 to 40 \\
Index ranges & $-8 \leq  h \leq  8, -14 \leq   k \leq   13,  -7 \leq  l \leq  7$\\
Reflections collected &	1866 \\
Independent reflections	& 467 [R$_{int}$ = 0.0727, R$_\sigma$ =  0.0468] \\
Data/restraints/parameters &	467/0/21 \\
Goodness-of-fit on F$^2$ & 1.049
 \\
Final R indexes [$I \ge	2\sigma(I)$] &	R$_1$ =  0.0337, wR$_2$ =  0.0780 \\
Final R indexes [all data] &	R$_1$ = 0.0341, wR$_2$ = 0.0782 \\
Largest diff. peak/hole, e \AA$^{-3}$	& 3.02/-1.65 \\
\mr
\br
\end{tabular*}
\end{table}

\begin{table}
\scriptsize
\caption{\label{atomic positions} {Atomic coordinates and anisotropic displacement parameters}}
\begin{tabular} {@{}lllllllllll}
\br
Atom &	x	& y &	z  & U$_{11}$	&  U$_{22}$ &	 U$_{33}$	&  U$_{23}$  &  U$_{13}$&  U$_{12}$\\
\mr
Yb &	0.5 & 0.33364(4) &	0.5 &	 0.0338(4) & 0.0227(3) & 0.0513(4) & 0.000 & 0.0192(3) & 0.000\\
Cl01&	0.7149(5)	& 0.5 &	0.7469(6) &	0.0440(18) & 0.0273(16) & 0.050(2) & 0.000 & 0.0072(15) & 0.000 \\
Cl02&	0.2413(4) & 0.1797(2) &	0.2603(5)	& 0.0446(13) & 0.0391(13) & 0.0508(14) & -0.0114(10) & 0.0244(11) & -0.0109(9) \\
\mr
\br
\end{tabular}
\end{table}

Layered van-der-Waals compounds are prone to stacking faults, often manifesting themselves in extensive diffuse streaks in their diffraction image, see, e.g. \cite{RuCl3_Johnson,RuCl3_Geck_arxiv}. 
As the reconstructed reciprocal space layers of our YbCl$_3$ crystals evidence, Fig. \ref{fig:h0l}, the Bragg reflections are not sharp spots but are slightly extended along [001] and, to a smaller extend, also along [100]. This indicates a certain amount of layer misorientation and domain formation but extensive stacking faults seem to be absent.

For the ESR experiment a millimetre sized single crystal was fixed with Apiezon grease in a suprasil quartz tube, which was sealed under vacuum. The crystal was oriented such that the microwave magnetic field direction is in the crystals' ab-plane. The crystal was then goniometer-aligned to the direction of the external field $B$ for angles between in-plane $B\parallel$~ab and out-of-plane $B\perp$~ab orientations. 

\section{Results}
%
%
%%%%%%%%%%%%%%%%%%%%%%%%%%%%%%%%%%%%%%%%%%%%%%%%
\begin{figure}[hbt]
\begin{center}
\includegraphics[width=0.8\columnwidth]{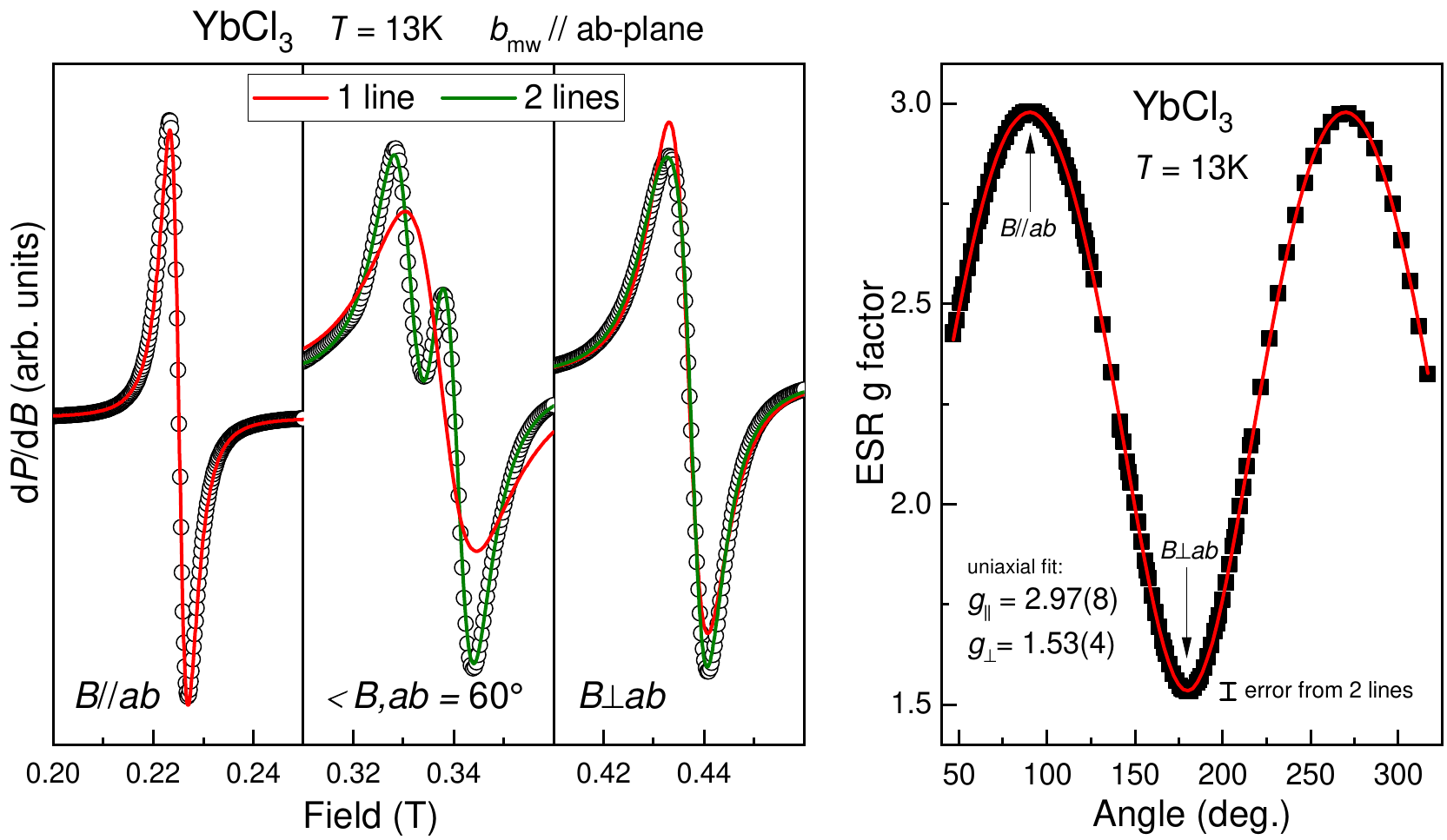}
\end{center}
\caption{Typical ESR spectra of YbCl$_{3}$ and their anisotropy at $T=13$~K. Left frame: ESR spectra for three crystal-$ab$-plane orientations in the external field $B$ and in the microwave field $b_{mw}$ as indicated. Solid lines denote Lorentzian lineshapes. For out-of-$ab$ plane field orientations two superimposed Lorentzian lines yield the best fit. Right frame: Anisotropy of the ESR $g$ factor resulting from fitting the spectra with one line. Dashed line indicates $g(\Theta) = \sqrt{g_\|^2\sin^2\Theta + g_\bot^2\cos^2\Theta}$ with $g_\|=2.97(8)$ and $g_\bot =1.53(4)$. The sample was rotated around an axis lying in the $ab$-plane parallel to $b_{mw}$. 
}
\label{Fig1}
\end{figure}
%%%%%%%%%%%%%%%%%%%%%%%%%%%%%%%%%%%%%%%%%%%%%%
%
Typical ESR spectra are shown in Fig. \ref{Fig1} for three $ab$-plane orientations in the external field $B$. Lorentzian line shapes were used to fit of the spectra (solid lines). 
%Noteworthy, for $B\perp ab$, the best data fit could be obtained by two superimposed Lorentzian lines which slightly differ in their $g$-value %$\Delta g^{\perp}= \pm 0.012$ $\Delta g^{\perp}= \pm 0.02$ and their linewidth $\delta\Delta B^{\perp}= \pm 1$~mT.
Satisfying fits could be achieved by a single line for in-plane field orientation ($B\|ab$, red line) whereas for out-of-plane field directions, two superimposed Lorentzian lines (green line) clearly yield a better fit. 
In fact, even two lines are not enough to perfectly fit the line shape. There are tiny contributions from further lines all having their individual orientation dependencies. To understand this multiline feature one may think of out-of-plane stacking faults. These are indicated by a broadened crystal field excitation as discussed for inelastic neutron diffraction data \cite{sala19a}. Such stacking faults result in a variation of the crystal field potential what in turn affects g-value and relaxation. However, as discussed above, the character of the Bragg reflections (Fig.  \ref{fig:h0l}) do not point toward extensive stacking faults but merely indicate a certain amount of domain formation and layer misorientation. The latter, in fact, provides the most sound explanation of the observed multi-lineshape. 
A two-line fitting results in g-value anisotropies which are shifted by $\approx 2^{\circ}$. Hence, dominantly two layers tilted by $\approx 2^{\circ}$ contribute to the lineshape being consistent with two superimposed lines.  

In order to extract the g-value anisotropy of the system as presented in Fig. \ref{Fig1}, right frame, we utilized a single line fitting (accepting a small additional error as indicated) and obtained for $T=13$~K (smallest linewidth) an in-plane $g^{\rm ESR}_\|=2.97(8)$ and an out-of-plane $g^{\rm ESR}_\bot =1.53(4)$. Note that these values display a much larger anisotropy than reported from low-temperature susceptibility data, $g^{\chi}_{\|}= 3.6(1)$ and  $g^{\chi}_{\perp} = 3.5(1)$ \cite{xing20b}. 
%The reason for this discrepancy is not obvious but could be related to the fact that ESR probes the Yb-magnetism locally and without any Van-Vleck contribution.\\
We suspect that this discrepancy is due to the incomplete analysis of the magnetization data available in the literature. In order to obtain a consistent description, high field magnetizations at low temperatures (far below the first crystal field level) have to be performed. At very high fields, the magnetization is linearly proportional to the magnetic field. The proportionality constant is the temperature independent and anisotropic Van Vleck susceptibility which must be considered as a contribution in the analysis of the susceptibility, especially in the determination of the magnetic moment (see \cite{ranjith19b}, for instance). 
%Without knowledge of the Van Vleck susceptibility, a determination of the moment in the spin 1/2 state of the Kramers ion Yb is always subject to errors.  

\begin{figure}[h]
\begin{center}
\includegraphics[width=0.6\columnwidth]{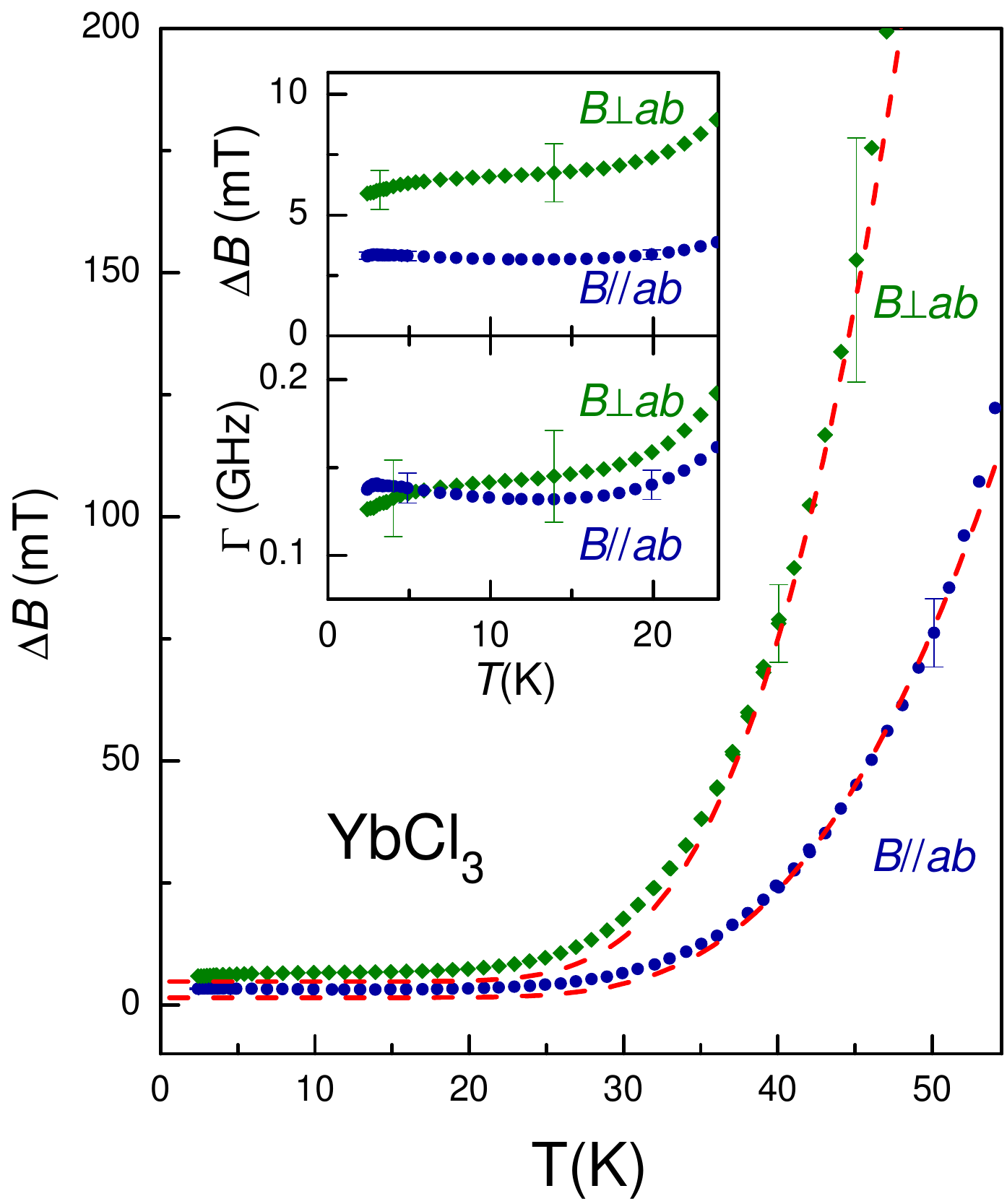}
\end{center}
\caption{
Temperature dependence of ESR linewidth $\Delta B$ in YbCl$_{3}$ for two orientations of the external field $B$ to the $ab$-plane. Dashed lines indicate towards higher temperatures the relaxation via the first excited crystalline electric field level of Yb$^{3+}$ at $\Delta=245\pm20$~K. Inset: Low-temperature close-up of linewidth and relaxation rate $\Gamma=\nu\Delta B/B_{\rm res}$.}
\label{Fig2}
\end{figure}

The temperature dependence of the linewidth $\Delta B(T)$ is shown in Fig. \ref{Fig2} and shows the same characteristics as found in Yb-based delafossites \cite{sichelschmidt19a,sichelschmidt20a}:
Towards high temperatures an Orbach process \cite{orbach61a,abragam70a} dominates the broadening according to $\Delta B(T) \propto 1/[\exp(\Delta/T)-1]$ with $\Delta$ denoting the first excited crystalline-electric field split electronic energy level. The red dashed line in Fig. \ref{Fig2} corresponds to $\Delta=245\pm20$~K ($21\pm2$~meV) in agreement with neutron scattering results for the lowest crystal field split level at $\hbar\omega = 21.04$~meV (244~K) \cite{sala19a}.
This elevated value of $\Delta$ is in accordance with the scenario of an effective spin-1/2 state for the low-temperature regime which is indicated by the recovery of  the spin-1/2 entropy of $R\ln2$ seen in specific heat results at $\approx8$~K \cite{xing20b}.
%
%%%%%%%%%%%%%%%%%%%%%%%%%%%%%%%%%%%%%%%%%%%%%%%%
%%%%%%%%%%%%%%%%%%%%%%%%%%%%%%%%%%%%%%%%%%%%%%
%%

Towards low temperatures the behavior of $\Delta B(T)$ is clearly different to that observed in the Yb-delafossites \cite{sichelschmidt20a} (see inset top frame Fig. \ref{Fig2}). For the latter a growing influence of spin correlations lead to an increase of $\Delta B(T)$. In YbCl$_{3}$, however, a slight increase could only be detected for $B\| ab$ whereas for $B\perp ab$ $\Delta B(T)$ continues to decrease towards low temperatures. Regarding the effect of spin correlations this observation is remarkable in the presence of N\'eel-type antiferromagnetic order below 0.6~K \cite{hao20a,xing20b} (with ordered moments along the a-axis). However, as shown in the lower inset frame of Fig.\ref{Fig2}, the spin relaxation rate shows very weak anisotropy and can even be considered isotropic within experimental accuracy.

\begin{figure}[h]
\begin{center}
\includegraphics[width=0.6\columnwidth]{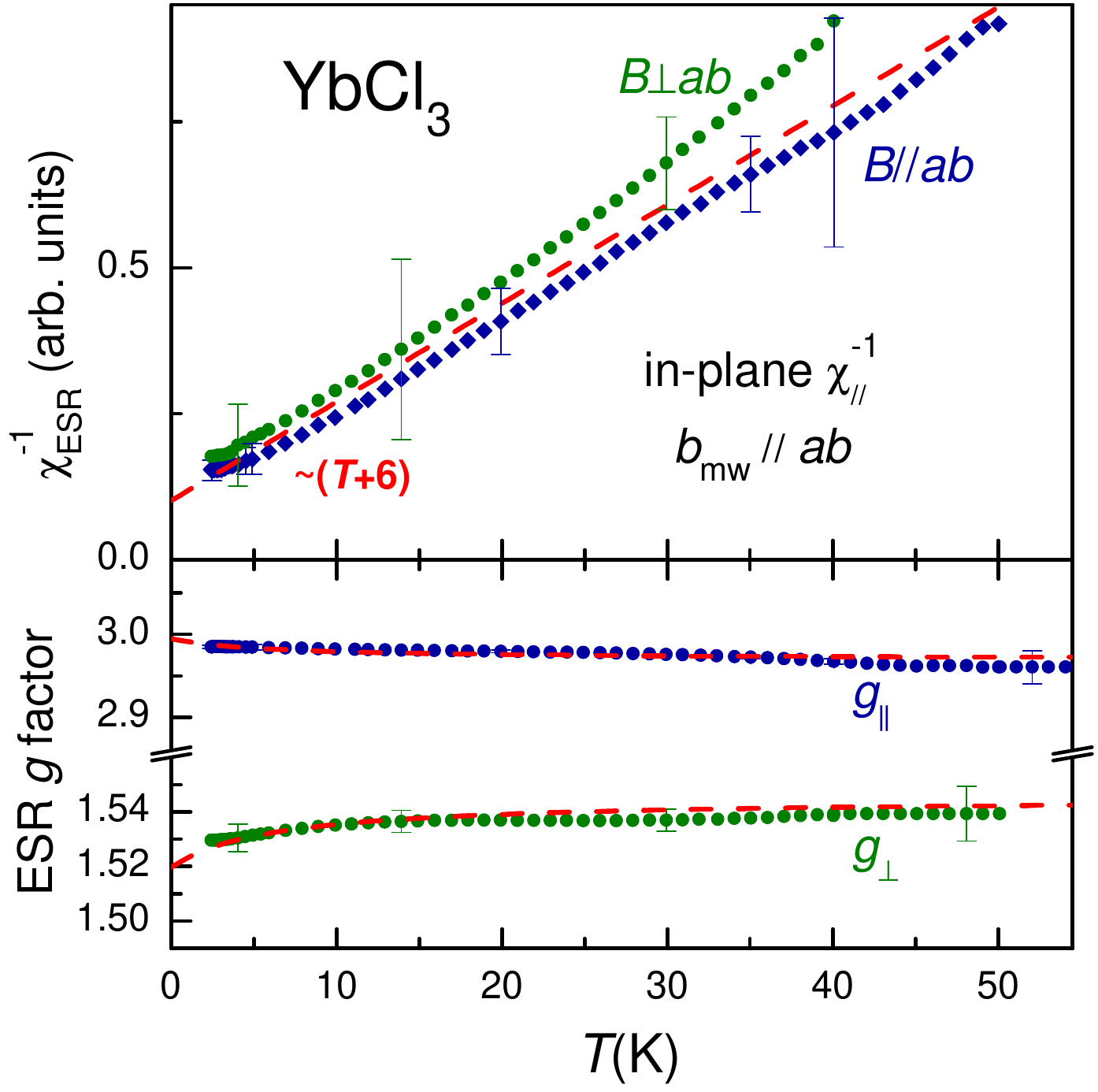}
\end{center}
\caption{
Temperature dependence of reciprocal ESR intensity $\chi^{-1}_{\rm ESR}$ and $g$ factor in YbCl$_{3}$ for the external field $B$ and the microwave field $b_{mw}$ aligned to the $ab$-plane as indicated. 
Dashed lines denote Curie-Weiss fits for the intensity and fits of the $g$-factor according Eqs.~(\ref{eq:1}) and (\ref{eq:2}), respectively.
}
\label{Fig3}
\end{figure}

The ESR intensity $I_{\rm ESR}\equiv \chi_{\rm ESR}$ is shown in the upper frame of Fig.~\ref{Fig3} for the microwave magnetic field $b_{mw}$ aligned in the $ab$-plane and for $B$ in-plane and out-of-plane directions, both yielding $\chi_{\|}$. Within the experimental error both dataset can be considered to be the same, as in the case of NaYbS$_{2}$ \cite{sichelschmidt19a}. Although the $B\perp ab$ intensity data are based on a single line fitting (see Fig.~\ref{Fig1}) the systematically larger $\chi_{\|}^{-1}$ for $B\perp ab$ cannot be explained if the intensity was based on a two lines fitting. Hence, both dataset can be reasonably well described by a Curie-Weiss law $\chi_{\|}^{-1}\propto T+\theta_{\|}$ with $\theta_{\|}=6$~K (solid line) which is consistent with the one from magnetization \cite{xing20b}, confirming dominant antiferromagnetic correlations in the plane.
%
% g factor

The ESR $g$ factor shows a weak temperature dependence, as shown in the lower frame of Fig.~\ref{Fig3}. There the dashed lines refer to a molecular magnetic field description of the anisotropic Yb-Yb interaction \cite{gruner10a,huber09a,sichelschmidt15a} providing a link between the $g$-factor and the exchange anisotropy which is reflected in $\theta_\bot - \theta_\|$ :
\begin{eqnarray}
\label{eq:1}
g_\|(T) &=& g_{\|}^0 \cdot \left(1-\frac{\theta_\| - \theta_\bot}{T + \theta_\|}\right)^\frac{1}{2}\\
\label{eq:2}
g_\bot(T) &=& g_{\bot}^0 \cdot \left(1+\frac{\theta_\| - \theta_\bot}{T + \theta_\bot}\right)
\end{eqnarray}

These equations provide a reasonable data description with $\theta_{\|}=6$~K given from $\chi_{\rm ESR}$. We obtained $g_{\bot}^0 \approx 2.97$, $g_{\|}^0 \approx 1.545$, and $\theta_\bot=6.1$~K for the adjustable parameters in good agreement with the $g$-values describing the anisotropy in Fig.\ref{Fig1}. Hence, the weak temperature dependence of $g(T)$ demonstrates a small difference $\theta_\bot - \theta_\|$, as similarly found for NaYbS$_{2}$ and NaYbSe$_{2}$ \cite{sichelschmidt20a}.

%%%%%%%%%%%%%%%%%%%%%%%%%%%%%%%%%%%%%%%%%%%%%%%
%%%%%%%%%%%%%%%%%%%%%%%%%%%%%%%%%%%%%%%%%%%%%
%

\section{Discussion and Summary}

The presented ESR studies provide the first results for local magnetic properties of Yb$^{3+}$ in YbCl$_{3}$ being partly dissimilar with non-local probes such as AC and DC susceptibility \cite{xing20b,hao20a,sala19a}.
The strong anisotropy of the $g$ factor (Fig.~\ref{Fig1}) as well as the exponential linewidth increase  (Fig.~\ref{Fig2}) are clear effects of the crystalline electric field that is locally acting on the Yb$^{3+}$ moments. These are centered in edge-sharing, tilted YbCl$_{6}$ octahedra 
giving rise to quasi-2D honeycomb layers of Yb$^{3+}$ ions. For this environment the low- temperature $J_{\rm eff}=1/2$ state of the Yb$^{3+}$ ion is characterized by the measured values in-plane $g_{\|}$ and out-of-plane $g_{\bot}$ given in Fig.~\ref{Fig1}. 
%
%Linewidth: Discussion of linewidth anisotropy: consistent with theta-difference?
As discussed for NaYbS$_{2}$ \cite{sichelschmidt19a} the small difference $\theta_\bot - \theta_\|$ in the Weiss temperatures obtained from the $g$ factor temperature dependence (Fig. ~\ref{Fig3}) is a measure of the anisotropy in the exchange interaction $J_{\mathrm{aniso}}$ between the Yb$^{3+}$ spins. This anisotropy leads to a broadening of the line in contrast to the isotropic exchange $J_{\mathrm{iso}}$ which is responsible for the exchange narrowing mechanism \cite{anderson53a}. With $\theta_\bot=6.1$~K, $\theta_\|=6$~K and with the notation and values used in Refs. \cite{sala19a,li15a} we obtained $J_{\mathrm{iso}}=(4J^{+-}+J^{zz})/3=4$~K.  
A rough estimate for the anisotropy broadening according to $\Delta B^{\mathrm{aniso}}\propto J_{\mathrm{aniso}}^{2}/J_{\mathrm{iso}}$ with $J_{\mathrm{aniso}}\propto|\theta_\|-\theta_\bot |=0.1$~K yields $\approx 1.2$~mT. This value is not small compared to the maximal estimations for hyperfine- ($\Delta B_{\rm h}=1$~mT) and dipolar broadening ($\Delta B_{\rm d}=0.04$~mT) using the nearest-neighbour Yb-distance of YbCl$_{3}$ \cite{sala19a}. Hence, in YbCl$_{3}$, similar to NaYbS$_{2}$ \cite{sichelschmidt19a}, the observed linewidth, reaching the smallest value of 3.16~mT for $B\| ab$-plane at 13~K, is largely due to a broadening from anisotropic exchange interactions.

In summary, YbCl$_{3}$ is an interesting quantum magnet. It remains to be seen whether other local probes such as muon spectroscopy or nuclear magnetic resonance will confirm the anisotropy determined here. It also remains to be seen whether the magnetic order at 0.6 K for YbCl$_{3}$ can be confirmed with these local probes. The related system YbBr$_{3}$, for example, shows an absence of magnetic order down to 0.1 K and is therefore classified as spin liquid \cite{wessler20a}.

\section*{Acknowledgements} We thank Prof. Anna Isaeva, U. Amsterdam, and Oliver Stockert for valuable discussions. Financial support by the Deutsche Forschungsgemeinschaft through SFB 1143 (project-id 247310070) is gratefully acknowledged.
\section*{References}

%\bibliography{JoergBib}

\begin{thebibliography}{10}
\expandafter\ifx\csname url\endcsname\relax
  \def\url#1{{\tt #1}}\fi
\expandafter\ifx\csname urlprefix\endcsname\relax\def\urlprefix{URL }\fi
\providecommand{\eprint}[2][]{\url{#2}}
% Bibliography created with iopart-num v2.1
% /biblio/bibtex/contrib/iopart-num

\bibitem{sichelschmidt19a}
Sichelschmidt J, Schlender P, Schmidt B, Baenitz M and Doert T 2019 {\em J.
  Phys.: Condens. Matter\/} {\bf 31} 205601

\bibitem{sichelschmidt20a}
Sichelschmidt J, Schmidt B, Schlender P, Khim S, Doert T and Baenitz M 2020
  {\em JPS Conf. Proc.\/} {\bf 30} 011096 (\textit{Preprint}
  \eprint{https://journals.jps.jp/doi/pdf/10.7566/JPSCP.30.011096})

\bibitem{schmidt21a}
Schmidt B, Sichelschmidt J, Ranjith K~M, Doert T and Baenitz M 2021 {\em Phys.
  Rev. B\/} {\bf 103}(21) 214445

\bibitem{haussler22a}
H\"au\ss{}ler E, Sichelschmidt J, Baenitz M, Andrade E~C, Vojta M and Doert T
  2022 {\em Phys. Rev. Materials\/} {\bf 6} 046201

\bibitem{trebst22a}
Trebst S and Hickey C 2022 {\em Phys. Rep.\/} {\bf 950} 1--37

\bibitem{rau18a}
Rau J~G and Gingras M~J~P 2018 {\em Phys. Rev. B\/} {\bf 98} 054408

\bibitem{xing20b}
Xing J, Feng E, Liu Y, Emmanouilidou E, Hu C, Liu J, Graf D, Ramirez A~P, Chen
  G, Cao H and Ni N 2020 {\em Phys. Rev. B\/} {\bf 102}(1) 014427

\bibitem{templeton54a}
Templeton D~H and Carter G~F 1954 {\em J. Phys. Chem.\/} {\bf 58} 940--944

\bibitem{hao20a}
Hao Y, Wo H, Gu Y, Zhang X, Gu Y, Zheng S, Zhao Y, Xu G, Lynn J~W, Nakajima K,
  Murai N, Wang W and Zhao J 2020 {\em Sci. China Phys.\/} {\bf 64} 237411

\bibitem{luo20a}
Luo Z~X and Chen G 2020 {\em SciPost Phys. Core\/} {\bf 3} 4

\bibitem{gruner10a}
Gruner T, Wykhoff J, Sichelschmidt J, Krellner C, Geibel C and Steglich F 2010
  {\em J. Phys.: Condens. Matter\/} {\bf 22} 135602

\bibitem{X-Shape}
Stoe \& Cie GmbH Darmstadt, Germany 2009 {\em \textsc{X-SHAPE}: Crystal
  Optimisation for Numerical Absorption Correction Program (Version 2.12.2)\/}

\bibitem{X-RED32}
Stoe \& Cie GmbH Darmstadt, Germany 2009 {\em \textsc{X-RED32}: Data Reduction
  Program (Version 1.53)\/}

\bibitem{shelxt}
Sheldrick G~M 2015 {\em Acta Crystallographica Section A\/} {\bf 71} 3--8

\bibitem{Olex2}
Dolomanov O~V, Bourhis L~J, Gildea R~J, Howard J~A~K and Puschmann H 2009 {\em
  Journal of Applied Crystallography\/} {\bf 42} 339--341

\bibitem{shelxl}
Sheldrick G~M 2015 {\em Acta Crystallographica Section C\/} {\bf 71} 3--8

\bibitem{RuCl3_Johnson}
Johnson R~D, Williams S~C, Haghighirad A~A, Singleton J, Zapf V, Manuel P,
  Mazin I~I, Li Y, Jeschke H~O, Valent\'{\i} R and Coldea R 2015 {\em Phys.
  Rev. B\/} {\bf 92}(23) 235119

\bibitem{RuCl3_Geck_arxiv}
Stahl Q, Ritschel T, Garbarino G, Cova F, Isaeva A, Doert T and Geck J 2022
  Pressure-tuning of $\alpha-\mathrm{RuCl}_{3}$ towards the ideal kitaev-limit
  arXiv:2209.08367

\bibitem{sala19a}
Sala G, Stone M~B, Rai B~K, May A~F, Parker D~S, Hal\'asz G~B, Cheng Y~Q,
  Ehlers G, Garlea V~O, Zhang Q, Lumsden M~D and Christianson A~D 2019 {\em
  Phys. Rev. B\/} {\bf 100}(18) 180406

\bibitem{ranjith19b}
Ranjith K~M, Luther S, Reimann T, Schmidt B, Schlender P, Sichelschmidt J,
  Yasuoka H, Strydom A~M, Skourski Y, Wosnitza J, K\"uhne H, Doert T and
  Baenitz M 2019 {\em Phys. Rev. B\/} {\bf 100} 224417

\bibitem{orbach61a}
Orbach R 1961 {\em Proc. Roy. Phys. Soc. A\/} {\bf 264} 458--484

\bibitem{abragam70a}
Abragam A and Bleaney B 1970 {\em Electron Paramagnetic Resonance of Transition
  Ions\/} (Oxford: Clarendon Press)

\bibitem{huber09a}
Huber D~L 2009 {\em J. Phys.: Condens. Matter\/} {\bf 21} 322203

\bibitem{sichelschmidt15a}
Sichelschmidt J, Gruner T, Jang D, Steppke A, Brando M, Mitsumoto K and Geibel
  C 2015 {\em J. Phys. Conf. Ser.\/} {\bf 592} 012017

\bibitem{anderson53a}
Anderson P~W and Weiss P~R 1953 {\em Rev. Mod. Phys.\/} {\bf 25} 269

\bibitem{li15a}
Li Y, Chen G, Tong W, Pi L, Liu J, Yang Z, Wang X and Zhang Q 2015 {\em Phys.
  Rev. Lett.\/} {\bf 115}(16) 167203

\bibitem{wessler20a}
Wessler C, Roessli B, Kr{\"a}mer K~W, Delley B, Waldmann O, Keller L,
  Cheptiakov D, Braun H~B and Kenzelmann M 2020 {\em npj Quantum Mater.\/} {\bf
  5} 85

\end{thebibliography}

\end{document}